\documentclass[a4paper,11pt]{article}
\usepackage{jheppub, bm, color} 
\usepackage{amssymb,amsfonts,slashed,amsthm,amsmath,graphicx, soul}
\usepackage[caption=false]{subfig}
\usepackage{slashed}

\renewcommand{\thefootnote}{\fnsymbol{footnote}}

\newcommand{\bea}{\begin{array}}
\newcommand{\eea}{\end{array}}
\newcommand{\beq}{\begin{eqnarray}}
\newcommand{\eeq}{\end{eqnarray}}

\newcommand{\MeV}{\  {\rm MeV} }
\newcommand{\GeV}{\  {\rm GeV} }
\newcommand{\TeV}{\  {\rm TeV} }

\newcommand{\lmk}{\left(}  
\newcommand{\rmk}{\right)}
\newcommand{\lkk}{\left[}  
\newcommand{\rkk}{\right]}

\newcommand{\Mpl}{M_{\rm Pl}}

\newcommand{\abs}[1]{\left\vert {#1} \right\vert}

\def\eq#1{Eq.~(\ref{#1})}

\definecolor{orange}{RGB}{255,100,0}
\definecolor{rosepink}{RGB}{248,100,100}

\begin{document}

\begin{flushright}
MIT-CTP/5149
\end{flushright}

\title{
A more attractive scheme for radion stabilization and supercooled phase transition
}

\author{
Kohei Fujikura,$^{1}$\footnote{
E-mail address: fuji@th.phys.titech.ac.jp} \ 
Yuichiro Nakai,$^{2}$\footnote{
E-mail address: ynakai@sjtu.edu.cn} \ and 
Masaki Yamada$^{3}\footnote{
E-mail address: yamada02@mit.edu}$\\*[20pt]
$^1${\it \normalsize
Department of Physics, Tokyo Institute of Technology,
Tokyo 152-8551, Japan} \\*[5pt]
$^2${\it \normalsize 
Tsung-Dao Lee Institute and School of Physics and Astronomy,\\ Shanghai
Jiao Tong University, 800 Dongchuan Road, Shanghai 200240, China} \\*[5pt]
$^3${\it \normalsize
Center for Theoretical Physics, Laboratory for Nuclear Science and Department of Physics, 
\\
Massachusetts Institute of Technology, Cambridge, MA 02139, USA}  \\*[50pt]
}

\abstract{
We propose a new radion stabilization mechanism in the Randall-Sundrum spacetime, introducing a bulk ${SU(N_H)}$ gauge field which confines at a TeV scale. It turns out that the radion is stabilized by the balance between a brane tension and a pressure due to the Casimir energy of the strong ${SU(N_H)}$ gauge field. We investigate the phase transition between the Randall-Sundrum (compactified) spacetime and a de-compactified spacetime and determine the parameter regime in which eternal (old) inflation is avoided and the phase transition can be completed. In comparison to the Goldberger-Wise mechanism, the 5D Planck mass can be larger than the AdS curvature and a classical description of the gravity is reliable in our stabilization mechanism. We also discuss the effect of the phase transition in cosmology such as an entropy dilution and a production of gravitational waves. 
}

\maketitle

\renewcommand{\thefootnote}{\arabic{footnote}}
\setcounter{footnote}{0}

%%%%%%%%%%%%%%%%%%%%%%%%%%%%%%%%%%%%%%%%%%%%%%%%%%%
\section{Introduction}
%%%%%%%%%%%%%%%%%%%%%%%%%%%%%%%%%%%%%%%%%%%%%%%%%%%

Results from collider experiments, including the discovery of the Standard Model (SM)-like Higgs boson, 
strongly indicate that the SM can explain phenomenologies around and below the electroweak scale.
However, the scale of the electroweak symmetry breaking (EWSB) is sensitive to high-energy physics, such as the grand unified theory (GUT) and the Planck scale physics, through radiative corrections.
The SM requires an {\it unnatural fine-tuning} to the Higgs potential to realize the correct EWSB. 
This is known as the hierarchy problem and has motivated various possibilities of physics beyond the SM. 

The Randall-Sundrum (RS) Model~\cite{Randall:1999ee} is an attractive scenario which 
provides an elegant solution to the hierarchy problem, introducing a warped extra dimension.
The geometry of the RS model consists of ${\rm AdS_5}$ bulk spacetime with two branes (called IR and UV branes) placed on two boundaries of the $5$-dimensional AdS bulk.
The SM Higgs field is assumed to be localized on the IR brane, while the massless graviton is localized toward the UV brane. 
An energy scale on the IR brane is exponentially redshifted from that on the UV brane and hence the hierarchy between the electroweak scale and the Planck scale is dynamically generated. 
According to the AdS/CFT correspondence \cite{Maldacena:1997re,Gubser:1998bc,Witten:1998qj,ArkaniHamed:2000ds,Rattazzi:2000hs}, the RS model is dual to a nearly-conformal strongly-coupled 4D field theory.
In the dual picture, the Higgs field is given as a bound state of this 4D theory.

The original RS model has
a massless modulus field called {\it radion}, which parameterizes the distance between the IR and UV branes,
and its vacuum expectation value is fixed by hand to realize an adequate redshift factor.
In the dual 4D picture, dilation invariance of the corresponding CFT is spontaneously broken and a massless Nambu-Goldstone boson
called dilaton exists.
To solve the (large) hierarchy problem completely, we thus need a mechanism that
stabilizes the radion vacuum expectation value (VEV) without fine-tuning. 
Many schemes for radion stabilization have been proposed so far~\cite{Goldberger:1999uk,Garriga:2000jb,Goldberger:2000dv,Hofmann:2000cj,Brevik:2000vt,Flachi:2001pq,Nojiri:2001ai,Garriga:2002vf,Haba:2019zjc},
including the famous Goldberger-Wise mechanism~\cite{Goldberger:1999uk} which introduces a bulk scalar field
with brane-localized potentials.

If the RS model is realized in nature, it must predict a consistent cosmological history of our Universe. 
At low temperature, the Universe is described by the compact RS model.
On the other hand, at high temperature,
the system is known to be described by the de-compactified AdS-Schwarzschild (AdS-S) solution
with the IR brane replaced by an event horizon~\cite{Creminelli:2001th} 
because the canonical ensemble of the AdS space is described by the AdS-S solution 
as argued by Hawking and Page \cite{Hawking:1982dh}. 
Therefore, as the temperature of the Universe cools down, 
a phase transition between the AdS-S spacetime and the RS spacetime takes place~\cite{Randall:2006py,Konstandin:2011dr}. 
Its 4D dual description is given by a confinement-deconfinement phase transition in the strongly-coupled 4D theory. 
It has been known that this phase transition is of the first order and proceeds via nucleation of true vacuum bubbles.

The phase transition between the AdS-S spacetime and the RS spacetime generally takes place
via a {\it supercooling} phase.\footnote{
Supercooling also takes place in other models such as singlet extensions of the SM and Coleman-Weinberg models (see Refs.~\cite{Witten:1980ez,Espinosa:2008kw,Jaeckel:2016jlh,Marzola:2017jzl,Iso:2017uuu,Brdar:2018num,Hashino:2018wee}).} 
This is easily understood in terms of the 4D dual picture: scale invariance of the CFT suppresses the phase transition
\cite{Buchmuller:1990ds}.
In particular, when we assume the Goldberger-Wise mechanism for radion stabilization, 
the supercooling phase lasts very long 
and the phase transition is never completed (and leads to eternal inflation)
in most of the region where the 5D Planck mass is much larger than the AdS curvature and
a classical treatment of the gravity is meaningful \cite{Creminelli:2001th}. 
Furthermore, even in the remaining parameter space,
the brane-localized potentials of the bulk scalar field give a non-negligible back-reaction to the gravitational action
and the analysis without including the back-reaction is not trustable.
Possible solutions to this problem have been discussed by several authors.
Refs.~\cite{Bunk:2017fic,Megias:2018sxv} explored soft wall models.
Ref.~\cite{Hassanain:2007js} considered a different geometry from the RS spacetime.
Ref.~\cite{Konstandin:2010cd} partially took into account the back-reaction in the Goldberger-Wise mechanism.
Ref.~\cite{Dillon:2017ctw} introduced a brane-localized curvature and made the phase transition faster. 
Furthermore, Refs.~\cite{vonHarling:2017yew,Baratella:2018pxi} took into account QCD effects on the radion potential  
and discussed that the phase transition is completed around the QCD dynamical scale.
Ref.~\cite{Agashe:2019lhy} constructed a dual 4D model having two renormalization fixed points which can make the phase transition faster.

In this paper, we propose a new mechanism of radion stabilization 
in which there is no issue in completion of the phase transition from the beginning, contrary to the Goldberger-Wise mechanism.
We introduce a hidden ${SU(N_H)}$ gauge field into the bulk of the extra dimension  
and assume that its asymptotically-free gauge coupling becomes strong and the theory confines at a TeV scale.\footnote{
The authors of Ref.~\cite{Luty:2000ec} have introduced a brane-localized Yang-Mills gauge field as well as
a bulk Yang-Mills field for radion stabilization in
the supersymmetric RS model while only a bulk Yang-Mills gauge field is introduced in our non-supersymmetric model.}
The confinement generates a vacuum energy that 
results in a pressure due to the Casimir force.\footnote{
There have been several studies of the Casimir energy in the extra dimension~\cite{Appelquist:1983vs,Appelquist:1982zs,Toms:2000vm} and challenges to stabilize the radion via the Casimir energy~\cite{Garriga:2000jb,Goldberger:2000dv,Hofmann:2000cj,Brevik:2000vt,Flachi:2001pq,Nojiri:2001ai}.}
The radion can be stabilized by the balance between the pressure due to the Casimir energy and the tension of the IR brane. 
The scale of the radion VEV is determined by the confinement scale of the hidden gauge field, 
which can be naturally of order the TeV scale, much smaller than the Planck scale, due to dimensional transmutation. 
Therefore, the electroweak naturalness is addressed without fine-tuning. 
As we will see, the phase transition can be completed
even when the 5D Planck scale is much larger than the AdS curvature scale. 
This justifies a classical treatment of the gravity in our analysis. 
In addition, since the $SU(N_H)$ gauge interaction is asymptotically free and 
the confinement scale is many orders of magnitude lower than the Planck scale, 
the Casimir energy is irrelevant at the Planck scale 
and any back-reaction from the gauge field to the gravitational action is trivially negligible. 

Although the phase transition can be completed in our mechanism of radion stabilization, there is still a supercooling phase. 
This fact leads to interesting cosmological phenomena such as an entropy dilution and a production of gravitational waves (GWs).
The strong supercooled phase transition potentially triggers the first-order electroweak phase transition which can be a promising candidate for the electroweak baryogenesis~\cite{Bruggisser:2018mus,Bruggisser:2018mrt}.
Moreover, a long supercooling epoch, which is characteristic to this kind of models, 
results in an almost maximal GW amplitude~\cite{Randall:2006py,Konstandin:2011dr}, which can be detected by future experiments such as eLISA~\cite{Seoane:2013qna}, DECIGO~\cite{Seto:2001qf} and BBO~\cite{Harry:2006fi}.

The rest of the paper is organized as follows.
In Sec.~\ref{sec:stabilization mechanism}, after a brief review of the RS model,
our new mechanism of radion stabilization is presented. 
In Sec.~\ref{sec:phase transition}, we consider thermal effects on the system and analyze the phase transition
between the AdS-S spacetime and the RS spacetime.
In Sec.~\ref{sec:supercooled universe}, we briefly discuss cosmological consequences of the phase transition
through a supercooling phase, especially focusing on generation of GWs. 
We also mention phenomenology of glueballs in our model.
Sec.~\ref{sec:conclusion} is devoted to conclusions.

\section{Radion stabilization mechanism} \label{sec:stabilization mechanism}

In this section, we first review basic properties of the RS model~\cite{Randall:1999ee}.
In particular, the effective action of the radion is summarized. 
Then, we explain our radion stabilization mechanism with a bulk Yang-Mills gauge field.
The radion mass is also calculated.

\subsection{The RS model}

The geometry of the RS spacetime is described by ${\bold R}^4 \times {\rm S}^1 / {\bold Z}_2$ 
with the following metric:
\begin{align}
ds^2 = G_{AB} dx^A dx^B = e^{-2kT(x)\left.| y \right.|} g_{\mu\nu} dx^\mu dx^\nu - T^2 (x) dy^2, 
\label{eq:RS metric}
\end{align}
where $A= (\mu, y)$ and the greek indices $\mu, \nu$ run from $0$ to $3$, 
$g_{\mu \nu}$ and  $k$  are the 4D induced metric and the AdS curvature of ${\cal O}(\Mpl)$, 
and $y$ $\in (-1/2,  1/2)$ represents the coordinate for the $5$th dimension.
We impose a ${\bold Z}_2$ symmetry $y \leftrightarrow -y$. 
Two $3$-branes, called UV and IR branes, 
are placed at the orbifold fixed points at $y=0$ and $y = y_{\rm IR} =1/2$, respectively. 
$T(x)$ determines the size of the extra dimension and 
is a modulus field associated with a fluctuation along the extra dimension. 
A pure gravitational action of the RS model is given by
\begin{align}
S = \int d^4 x dy \lkk \sqrt{G} \left( \frac{1}{2} M_5^3 R - \Lambda_{\rm bulk} \right)
- \Lambda_{\rm IR}  \sqrt{-g_{\rm IR}} \, \delta (y-y_{\rm IR})  - \Lambda_{\rm UV} \sqrt{-g_{\rm UV}} \, \delta (y) \rkk, 
\label{eq:E-H action}
\end{align}
where we take the domain of the integral for the 5th dimension to be $(-1/2 , 1/2)$. 
Here, $M_5$ and $R$ are the 5D Planck mass and the Ricci scalar, 
and $\sqrt{G},~\sqrt{-g_{\rm IR}}$ and $\sqrt{-g_{\rm UV}}$ represent the volume elements of the bulk metric and 
the induced metrics at the IR and UV branes, respectively.
In addition, $\Lambda_{\rm bulk}$ is a bulk cosmological constant, and
$\Lambda_{\rm IR}$ and $\Lambda_{\rm UV}$ are IR and UV brane tensions.
The geometry of \eqref{eq:RS metric} is realized when we tune the cosmological constant and the brane tensions as
$ \Lambda_{\rm bulk} |_{\rm RS} /k =  \Lambda_{\rm IR}|_{\rm RS} = - \Lambda_{\rm UV}|_{\rm RS} = - 6 M_5^3 k$.
From Eqs.~\eqref{eq:RS metric} and \eqref{eq:E-H action}, one can find that 
every mass parameter on the IR brane is suppressed by  
the warp factor $e^{-kT_0 /2}$ where $T_0$ is the modulus VEV, 
when measured with the 4D Einstein metric. 
On the other hand, since the 4D graviton wave-function is localized toward the UV brane,
the 4D Planck scale $M_{\rm Pl}$ is not strongly redshifted,
\begin{align}
M_{\rm Pl}^2  =  M_5^3 k^{-1} (1- e^{-kT_0}).
\end{align}
Therefore, the hierarchy problem is addressed if the SM Higgs field is localized on the IR brane and $k T_0 \approx 70$ is realized.
In general, we can consider a (small) deviation from the relation,
$ \Lambda_{\rm bulk} |_{\rm RS} /k =  \Lambda_{\rm IR}|_{\rm RS} = - \Lambda_{\rm UV}|_{\rm RS} = - 6 M_5^3 k$,
by shifting the brane tensions as 
\beq 
 \Lambda_{\rm IR} = - 6 M_5^3 k + \delta \Lambda_{\rm IR}, 
 \quad 
 \Lambda_{\rm UV} = 6 M_5^3 k + \delta  \Lambda_{\rm UV}.
\eeq
We include these shifts in the following discussions.

For a later use, let us consider the action of the modulus field $T(x)$.
The 4D effective action is derived from the Kaluza-Klein (KK) reduction of the pure gravitational action \eqref{eq:E-H action} and 
given by~\cite{Goldberger:1999un}
\begin{equation}
\begin{split}
S_{\rm eff} = \,&\frac{M_5^3}{2k} \int d^4x \sqrt{-g} \left(1-e^{-kT(x)}\right)R^{(4)} \\
&+ \frac{3}{4} M_5^3 k \int d^4x \sqrt{-g} \, \partial_\mu T(x) \partial^\mu T(x) e^{-kT(x)} \\
&+S_{\rm IR}+S_{\rm UV},
\label{modulusaction}
\end{split}
\end{equation}
where $R^{(4)}$ is the 4D Ricci scalar calculated by the induced metric $g_{\mu\nu}$ and  $S_{\rm IR}$ and $S_{\rm UV}$ are
defined as
\begin{align}
&S_{\rm IR } = -\int d^4 x \sqrt{-g}  \, e^{-2kT(x)} \delta \Lambda_{\rm IR},\\
&S_{\rm UV}= -\int d^4x\sqrt{-g}  \, \delta \Lambda_{\rm UV} .\label{eq:radion action}
\end{align}
We now define the radion field $\mu \equiv k e^{-k T(x)/2}$ by field redefinition from the modulus field $T(x)$.
The effective action of the radion field is then written as
\begin{align}
&S_{\rm radion} = \int d^4 x \left[ \, \frac{3N^2}{4\pi^2}  \left( \partial \mu(x)\right)^2 - V(\mu) \right], \label{radionkinetic} \\[1.5ex]
&V(\mu)= \delta \Lambda_{\rm UV} + \mu^4 \delta \Lambda_{\rm IR}/k^4, \label{eq:radion effective potential}
\end{align}
where we assume a flat spacetime in 4D for simplicity.\footnote{
The first line in Eq.~\eqref{modulusaction} contains a mixing term 
between the 4D scale factor and the radion field~\cite{Charmousis:1999rg}.
In fact, when we consider a non-trivial 4D background geometry such as an expanding Universe,
it is needed to diagonalize the kinetic term~\cite{Csaki:1999mp}. 
In this case, we have an additional factor in Eq.~\eqref{eq:radion effective potential}, but 
this factor gives a negligible contribution to the potential.}
Here, we have defined $N \equiv 2\pi (M_5 / k)^{3/2}$ and the radion kinetic term is not canonically normalized. 
We can recover the RS geometry by tuning the two shifts of the brane tensions to zero,
$\delta \Lambda_{\rm UV} = \delta \Lambda_{\rm IR} =0$. 
The one corresponds to the usual tuning of a vanishingly small cosmological constant at present. 
The other tuning is specific to the RS model and 
can be avoided if we can stabilize the radion $\mu$ at an appropriate value
by generating its potential from some mechanism on top of the one in \eqref{eq:radion effective potential}. 
As far as  $|\delta \Lambda_{\rm UV}|$ and $|\delta \Lambda_{\rm IR}|$ are small compared to
$6 M_5^3 k$,
their effects on the RS geometry are negligible.

Here we comment on the 4D dual picture of the RS model.
Our ${\rm AdS_5}$ bulk spacetime corresponds to a strongly interacting 4D CFT whose number of colors is
$N$ defined above \cite{ArkaniHamed:2000ds,Rattazzi:2000hs}. 
The presence of the IR brane corresponds to spontaneous breaking of the conformal symmetry at the scale
$\mu_0 \equiv k e^{-k T_0/2}$. 
Since we are interested in the regime of a large $M_5 / k$ where quantum gravity effects are neglected,
the number of colors $N$ should be large.
From the naive dimensional analysis~\cite{Agashe:2007zd}, 
terms with higher powers of the Ricci scalar coming from quantum gravity effects can be neglected for~\cite{vonHarling:2017yew} 
\beq
\label{Nmin}
 N \gtrsim 4 \cdot 5^{3/4} / \sqrt{3 \pi} \simeq 4.4. 
\eeq
We consider the case in which this condition is satisfied. 

\subsection{A new scheme for radion stabilization} \label{sec:stabilization}

We shall now provide our radion stabilization mechanism.
Let us introduce a $SU(N_H)$ pure Yang-Mills field that resides in the bulk of the extra dimension.\footnote{
Our discussion is similar to the case of the ordinary QCD that has been investigated in Ref.~\cite{vonHarling:2017yew}.
However, the point is that we utilize the radion potential generated by new strong dynamics to stabilize the radion while
we cannot expect such a large contribution to the radion potential in the ordinary QCD.}
We can introduce matter fields charged under the $SU(N_H)$ gauge group but they are irrelevant to our discussion.
The action for the gauge field is given by
\begin{equation}
\begin{split}
S_{\rm Yang-Mills}= \int d^5 x \left[ \sqrt{G} \left(-\frac{1}{4g_5^2} F_{AB} F^{AB} 
\right) -\sqrt{-g_{\rm IR}}\left(\delta(y-y_{\rm IR})\frac{\tau_{\rm IR}}{4}F_{\mu\nu}F^{\mu\nu}\right) \right.  \\  \left.
-\sqrt{-g_{\rm UV}}\left(\delta(y)\frac{\tau_{\rm UV}}{4}F_{\mu\nu}F^{\mu\nu}\right)\right],
\end{split}
\end{equation}
where $F_{AB}$ and $g_5$ are the 5D gauge field strength and the gauge coupling constant, while $F_{\mu\nu}$ is the 4D gauge field strength and $\tau_{\rm IR(UV)}$ parameterizes the IR (UV) brane localized kinetic term.
After the KK decomposition and integrating over the extra dimension,
we obtain the following 4D effective action for the zero-mode gauge field: 
\begin{align}
S_{\rm Yang-Mills}^{(0)} = \int d^4x \left[-\frac{1}{4}\left( 
\frac{\log\frac{k}{\mu}}{ kg_5^2} +\tau_{\rm IR}+\tau_{\rm UV}\right) 
F^{(0)}_{\mu\nu} F^{(0)\mu\nu}\right],
\end{align}
where $F^{(0)}_{\mu\nu}$ represents the field strength of the zero-mode.

Including the effect of the renormalization
group running from the UV scale $k$ to an energy scale $Q$,
one can express the 4D gauge coupling of the zero-mode gauge field $g_4(Q,\mu)$ as~\cite{Agashe:2002bx,Csaki:2007ns}
\begin{align}
\frac{1}{g_4^2 (Q,\mu)} = \frac{\log \frac{k}{\mu}}{k g_5^2}+\tau_{\rm IR}+\tau_{\rm UV}
-\frac{b_{\rm YM}}{8\pi^2} \log\left(\frac{k}{Q}\right) 
~~ {\rm for} ~~Q\lesssim\mu \, . \label{eq:matching condition}
\end{align}
The $\beta$-function coefficient is given by $b_{\rm YM}= 11 N_H /3 $.
In the dual 4D picture, we can understand the first term in the right hand side 
as 
the running factor due to the CFT degrees of freedom, 
which are confined at the scale of $\mu$ and are absent below that energy scale. 
It is then convenient to rewrite its prefactor as 
\begin{align}
\frac{1}{kg_5^2} =- \frac{b_{\rm CFT}}{8\pi^2}, 
\label{eq:CFT beta-function coefficient}
\end{align}
and we expect $b_{\rm CFT} = - \alpha N$ with $\alpha$ being a positive constant. 
The confinement scale of this gauge theory $\Lambda_H (\mu)$ is determined by
the condition $g_4^2 (Q\equiv \Lambda_{H}, \mu) = \infty$.
From Eqs.~(\ref{eq:matching condition}) and (\ref{eq:CFT beta-function coefficient}), 
we obtain 
\begin{equation}
\begin{split}
\Lambda_{H} (\mu) 
= \left( e^{-8\pi^2(\tau_{\rm IR}+\tau_{\rm UV})}\left(\frac{\mu}{k}\right)^{-b_{\rm CFT}} k^{b_{\rm YM}} 
\right)^{1/b_{\rm YM}} 
\equiv \Lambda_{H,0} \left( \frac{\mu}{\mu_{\rm min}}\right)^n,
\label{eq:radion dependent confinement scale}
\end{split}
\end{equation}
for $\Lambda_{H} (\mu) \lesssim \mu$, 
where 
$\mu_{\rm min}$ and 
$\Lambda_{H,0}$ are the radion VEV at the minimum of the potential specified later and the confinement scale at present, respectively. 
The (positive) exponent $n$ is defined by
\begin{align}
n\equiv -\frac{b_{\rm CFT}}{b_{\rm YM}} = \alpha\frac{3}{11}\frac{N}{N_H}. \label{eq:n}
\end{align}
For convenience, 
we introduce an ${\cal O}(1)$ unknown factor $n_c$ to parametrize our ignorance of the threshold between the confinement and deconfinement phases: 
\beq
 \Lambda_{H} (\mu_c)  \equiv \gamma_c \mu_c, 
\eeq
where $\mu_c$ is defined in such a way that \eq{eq:radion dependent confinement scale} is valid for $\mu_c \le \mu$. 
Note that the description of the 4D effective theory breaks down
when the confinement scale is larger than the lightest KK mass of the gauge field, $m_{KK}=\pi \mu$.
We thus expect $\Lambda_{H}(\mu_c) \simeq \pi \mu_c$, that is, $\gamma_c \simeq \pi$.

Next, let us consider the case for $\mu_c > \mu$, 
where the description of the 4D effective theory breaks down. 
In this case we can understand the behavior of the dynamical scale by the AdS/CFT correspondence. 
The CFT is not confined and contributes to the running to the gauge coupling until the $SU(N_H)$ gauge interaction becomes strong. 
Thus we expect that the confinement scale is independent of the radion VEV $\mu$ 
and obtain 
\begin{align}
\Lambda_{H} (\mu) 
= \Lambda_{H} (\mu_c) ~~(\equiv \gamma_c \mu_c)  ~~ {\rm for}~~ \mu < \mu_c \, . \label{eq:constant confinement scale}
\end{align}
Since this should be equal to \eq{eq:radion dependent confinement scale} at $\mu = \mu_c$, 
we can determine $\mu_c$ as 
\beq
 \mu_c = \mu_{\rm min} \lmk \frac{\Lambda_{H,0}}{\gamma_c \mu_{\rm min}} \rmk^{1/(1-n)}. 
 \label{muc}
\eeq

Equipped with the radion dependence of the confinement scale, we can discuss the radion potential generated by the confinement of
our gauge theory. 
First, we note that the trace of the energy momentum tensor for the $SU(N_H)$ gauge field is nonzero due to the conformal anomaly and is given by 
\begin{align}
 T^\mu_\mu = -\frac{b_{\rm YM}}{32\pi^2} F^{(0)}_{\mu\nu} F^{(0) \mu\nu}. 
\end{align}
The expectation value of the right hand side is the gluon condensate, 
which we expect 
\begin{align}
\langle F^{(0)}_{\mu\nu} F^{(0) \mu\nu} \rangle  \sim (4\pi)^2 \Lambda^4_{H} (\mu), \label{eq:hidden gluon condensation}
\end{align}
from the dimensional analysis.
Then, the vacuum energy is given by 
\begin{align}
 V_{H} = \frac{1}{4} \langle T_\mu^\mu \rangle \simeq -\frac{b_{\rm YM}}{8} \left(\Lambda_{H} (\mu) \right)^4. 
 \label{V_QCD'}
\end{align}
According to the lattice calculation,
the coefficient $1/8$ is replaced by $1/17$ for the case of the SM QCD~\cite{vonHarling:2017yew},
which supports our ${\cal O}(1)$ estimation. 
Combining with the potential \eqref{eq:radion effective potential}, we can summarize the total radion potential as follows: 
\begin{align}
V_{r,{\rm eff}} (\mu)=
\begin{cases}
V_0 + \frac{\lambda}{4} \mu^4 -\frac{b_{\rm YM}}{8} \Lambda^4_{H, 0} \left(\frac{\mu}{\mu_{\rm min}}\right)^{4n} &~~{\rm for} \  \mu > \mu_c \, ,\\
V_0 + \frac{\lambda}{4}  \mu^4 -\frac{b_{\rm YM}}{8} \gamma_c^4 \mu_c^4 &~~{\rm for} \  \mu< \mu_c \, , \label{eq:radion potential}
\end{cases}
\end{align}
where $\lambda \equiv 4 \delta \Lambda_{\rm IR} / k^4$ 
comes from the IR brane tension and we assume a positive $\lambda$.
Here $V_0 \equiv  \delta \Lambda_{\rm UV}$ is determined by the condition that the potential energy at the present vacuum $V(\mu_{\rm min})$ is vanishingly small as the observation of the dark energy indicates. 
The third terms come from \eq{V_QCD'} with $\Lambda_{H} (\mu)$ given by \eq{eq:radion dependent confinement scale} or \eq{eq:constant confinement scale}. 
We note that $n$ must be smaller than unity to stabilize the radion at a finite field value, 
since otherwise the potential has no minimum other than $\mu = 0$. 
With $n<1$, the radion VEV at the potential minimum is determined as
\begin{align}
\mu_{\rm min} = \left( \frac{n b_{\rm YM}}{2 \lambda}\right)^{\frac{1}{4}} \Lambda_{H, 0} \, . 
\label{mu_min}
\end{align}
We note that $\mu_{\rm min}$ must be larger than $\mu_c$,
which implies 
\begin{align}
\left(\frac{n b_{\rm YM}}{2 \lambda}\right)^{\frac{1}{4}} \gamma_c > 1 \,  . \label{ncondition1}
\end{align}
The potential energy at the minimum should be vanishingly small (except for a small cosmological constant), so that $V_0$ is determined by 
\beq
 V_{r, {\rm eff}} (\mu_{\rm min}) = V_0 - \frac{\lambda}{4} \lmk \frac{1-n}{n} \rmk \mu_{\rm min}^4 = 0. 
\eeq
From \eq{eq:radion dependent confinement scale} and $(\mu_c / \mu_{\rm min})^n < 1$, 
we can see that $\Lambda_{H} (\mu_c)$ is smaller than $\Lambda_{H, 0}$. 
Equation~\eqref{ncondition1} as well as $n<1$ constrain the number of colors $N_H$ of
the new Yang-Mills gauge theory for each $N$.

To calculate the radion mass, we note that the kinetic term of the radion $\mu$ in \eqref{radionkinetic} is not of the canonical form.
Canonically normalizing the kinetic term,
the physical mass of the radion at the potential minimum $\mu=\mu_{\rm min}$ is given by
\begin{align}
 m_{\rm radion}^2 = \left( \frac{2\pi^2}{3N^2}\right)
 4\lmk 1-n \rmk \lambda 
  \mu_{\rm min}^2 \, . \label{eq:radion mass}
\end{align}

%%%%%%%%%%%%%%%%%%%%%%%%%%%%%%%%%%%%%%%%%%%%%%%%%%%%
\begin{figure}[!t]
\begin{minipage}{0.32\hsize}
\begin{center}
\includegraphics[clip, width=4.5cm]{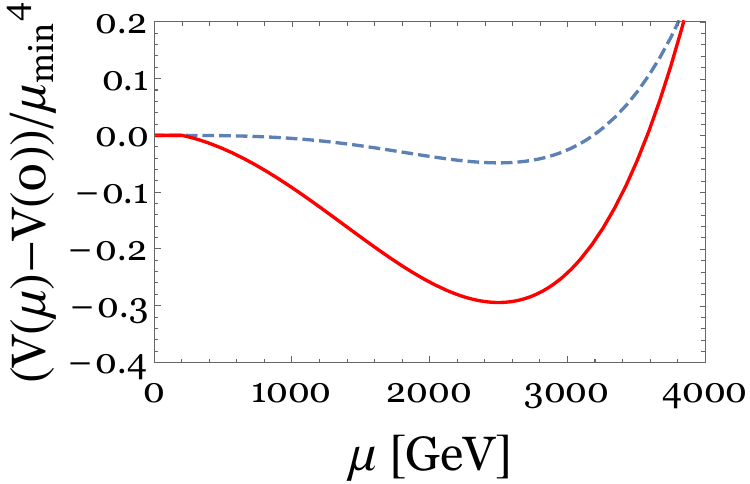}
\end{center}
\end{minipage}
\begin{minipage}{0.32\hsize}
\begin{center}
\includegraphics[clip, width=4.8cm]{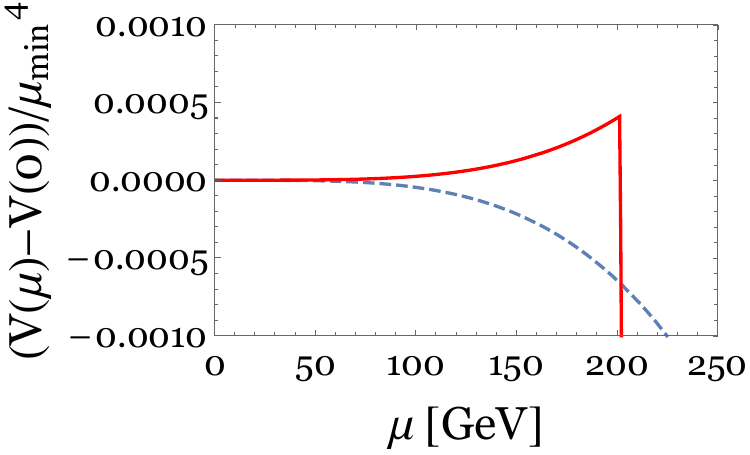}
\end{center}
\end{minipage}
\begin{minipage}{0.32\hsize}
\begin{center}
\includegraphics[clip, width=5.4cm]{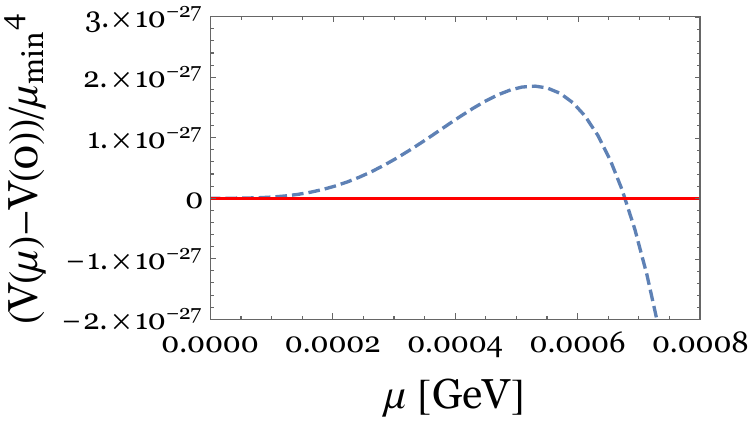}
\end{center}
\end{minipage}
\caption{The radion potential given by \eq{eq:radion potential} (red solid curve) 
for the case of $\mu_{\rm min} =2.5 \, {\rm TeV},~\lambda=1,~\gamma_c=\pi,~\alpha=1,~N=5$ and $N_H=3$ ($m_{\rm radion}\simeq 2.2 \, \rm TeV$). 
We also plot the radion potential calculated by the Goldberger-Wise mechanism (blue dotted curve), where we take parameters such that the radion VEV at the potential minimum is the same as ours. 
In both cases, the kinetic term is given by \eqref{radionkinetic}.
The left, middle and right panels focus on the regions around $\mu=\mu_{\rm min}$, $\mu =\mu_c$ and $\mu=0$,
respectively. 
}\label{fig1}
\end{figure}
%%%%%%%%%%%%%%%%%%%%%%%%%%%%%%%%%%%%%%%%%%%%%%%%%%%%

Figure~\ref{fig1} shows the radion potential in our model of radion stabilization (solid curve).
The kinetic term is given by \eqref{radionkinetic}. We take $\mu_{\rm min} =2.5 \, {\rm TeV},~\lambda=1,~\gamma_c=\pi,~\alpha=1,~N=5$ and $N_H=3$.
The radion mass is then estimated as $m_{\rm radion}\simeq 2.2 \, \rm TeV$.
Since the Casimir energy is constant for $\mu<\mu_c$, the potential due to the brane tension, $\lambda \mu^4 /4$, determines the shape of the potential around $\mu=0$. 
This implies that the origin of the potential is a local minimum as we can see from the middle panel of Fig.~\ref{fig1}. 
The Casimir energy becomes larger in magnitude for $\mu > \mu_c$ and dominates the potential. 
However, 
it is proportional to $\mu^{4n}$ and the potential due to the brane tension is proportional to $\mu^4$, so that the potential at a larger
radion VEV is dominated by the potential due to the brane tension. 
As a result, there is a minimum at $\mu = \mu_{\rm min}$ given by \eq{mu_min}. 
Note that $\mu_{\rm min}$ is roughly given by $\Lambda_{H, 0}$ 
because it is the typical energy scale in our stabilization mechanism. 
As a comparison, in figure~\ref{fig1}, we also show the radion potential for the case of the Goldberger-Wise mechanism,
which is given by Fig.~2 of Ref.~\cite{vonHarling:2017yew}. 
As can be seen from the left panel of Fig.~\ref{fig1},
the radion potential in our mechanism has a deeper minimum 
than the one in the Goldberger-Wise mechanism. 
There is also a local minimum at $\mu = 0$ in the case of the Goldberger-Wise mechanism~\cite{Goldberger:1999uk}. 
The first derivative of the radion potential in our mechanism is not continuous at $\mu = \mu_c$, reflecting our ignorance of
the precise radion potential around this point.
We will approximate the radion potential near $\mu = \mu_c$ by an analytic function to evaluate a tunneling action
of the transition from $\mu=0$ to $\mu=\mu_{\rm min}$ numerically.

\section{Phase transition from AdS-S to RS} \label{sec:phase transition}

In this section, we take account of thermal effects on the RS spacetime and discuss the phase transition between the AdS-S spacetime and the RS spacetime.
We determine the parameter region of our model where the phase transition is completed.

\subsection{Critical temperature and order parameter}

At high temperature, as argued in Ref.~\cite{Creminelli:2001th}, thermal corrections to the radion potential make the RS spacetime deform into the AdS-S spacetime with the IR brane replaced by the event horizon emitting the Hawking radiation.
As the temperature of the Universe cools down, the phase transition from the AdS-S spacetime to the RS spacetime can take place when the RS spacetime is energetically favored.

In order to clarify which spacetime is energetically favored, we first calculate the free energy of each spacetime.
The AdS-S spacetime is described by the following metric:
\begin{align}
ds^2 = k^2\rho^2 \left(1-\frac{\rho_H^4}{\rho^4} \right)dt^2 -  k^2 \rho^2  \sum^3_{i=1} dx_i^2
-\dfrac{d\rho^2}{k^2 \rho^2 \left(1-\frac{\rho_H^4}{\rho^4} \right) }, \label{eq:AdS-S}
\end {align}
where $\rho$ represents the coordinate for the 5th dimension. 
This metric covers $\rho_{\rm UV} > \rho > \rho_H$ where $\rho_{\rm UV}$ represents the position of the UV brane
and $\rho_H$ denotes the position of the event horizon.
The limit of $\rho_H =0 $ gives the bulk ${\rm AdS}$ metric:
\begin{align}
ds^2 = k^2 \rho^2 \left(dt^2 -\sum^3_{i=1}d x_i^2\right)-  \dfrac{1}{k^2\rho^2}d\rho^2 ,
\end{align}
which corresponds to the RS metric \eqref{eq:RS metric} with the identification of $\rho=k^{-1} \exp(-kT_0y/2)$,
taking $g_{\mu\nu}= {\rm diag} (1, -1, -1, -1)$.
The free energy of the AdS-S spacetime subtracted by that of the bulk AdS spacetime,
$\Delta F_{\rm AdS-S}$, is evaluated in Ref.~\cite{Creminelli:2001th} as
\begin{align}
\Delta F_{\rm AdS-S}(T_H) = \frac{3}{8}\pi^2 N^2 T_H^4 -\frac{1}{2} \pi^2 N^2 T_H^3 T , \label{eq:AdS-S free energy}
\end{align}
where $T_H$ ($\equiv k^2 \rho_H / \pi$) is the Hawking temperature parameterized by the position of the event horizon.
The minimum of this free energy is given by $T_H =T$.
Away from the minimum, a conical singularity appears at the event horizon in the Euclidean coordinate.
In the calculation of the free energy \eqref{eq:AdS-S free energy}, we only estimate the thermal contribution from the gravity part.
However, there is an additional contribution to the free energy from the bulk $SU(N_H)$ gauge field in our model.
We simply expect that this contribution is proportional to $N_H^2 T^4$,
but a coefficient is not determined since we need to evaluate it on the non-trivial AdS-S background.
For simplicity, we focus on the case where $N_H < N$ so that 
a thermal contribution from the gauge field is negligible.\footnote{In the case of the Goldberger-Wise mechanism, a thermal contribution from the bulk Goldberger-Wise field to the free energy is negligible as long as the back-reaction to the original RS spacetime is small~\cite{Creminelli:2001th}.}
We also note that $N$ must be larger than about $4.4$ from \eq{Nmin}, 
so that we expect that the free energy $\Delta F_{\rm AdS-S}(T_H)$ is large enough to dominate 
the thermal contributions from the SM particles.

The free energy of the RS spacetime (subtracted by that of the bulk AdS spacetime) is given by the radion potential at the minimum,
\begin{align}
\Delta F_{\rm RS} = V_{r,{\rm eff}} (\mu_{\rm min}) - V_{r,{\rm eff}} (0) , \label{eq:RS free energy}
\end{align}
where $V_{r,{\rm eff}}$ is defined in Eq.~\eqref{eq:radion potential}.
We have ignored the common constant in Eqs.~\eqref{eq:AdS-S free energy}, \eqref{eq:RS free energy}.
Since we use the 4D effective field theory to calculate the free energy, 
\eq{eq:RS free energy} is reliable only for $\mu \gtrsim T$. 
In this case, 
thermal contributions to $\Delta F_{\rm RS}$ can be neglected.

Let us now estimate the critical temperature of the phase transition between the AdS-S spacetime and the RS spacetime. 
The critical temperature $T_c$ is defined as the temperature when the free energies of the two phases are degenerated,
$\Delta F_{\rm AdS-S} (T) - \Delta F_{\rm RS}=0$.
From this condition, it is explicitly estimated as
\begin{align}
T_c = \left( 8\frac{V_{r, {\rm eff}} (\mu_{\rm min})}{\pi^2 N^2}\right)^{1/4}. \label{eq:critical temperature}
\end{align}
We can easily see from the above expression that there is no phase transition in the absence of a radion stabilization mechanism, $V_{r,{\rm eff}}=0$.
This fact can be easily understood from the dual perspective:
if the scale invariance was not explicitly broken in the confinement phase,
there would be no dimensionful parameter except the temperature in the theory~\cite{Witten:1998zw,Nardini:2007me}
and hence the system would be in the false vacuum forever no matter how small the temperature is.
When we introduce a radion potential to stabilize it, 
there is an explicit breaking for the scale invariance and the phase transition can take place.
As we will discuss below, in the case of the Goldberger-Wise mechanism,
the radion potential is nearly scale invariant, and hence, the phase transition is generally very slow.
On the other hand, in our stabilization mechanism, 
the strong dynamics of $SU(N_H)$ breaks scale invariant more strongly 
and the phase transition can be completed faster.

%%%%%%%%%%%%%%%%%%%%%%%%%%%%%%%%%%%%%%%%%%%%%%%%%%%%%%%%%%%%%
\begin{figure}[t]
\centering\includegraphics[width=8cm]{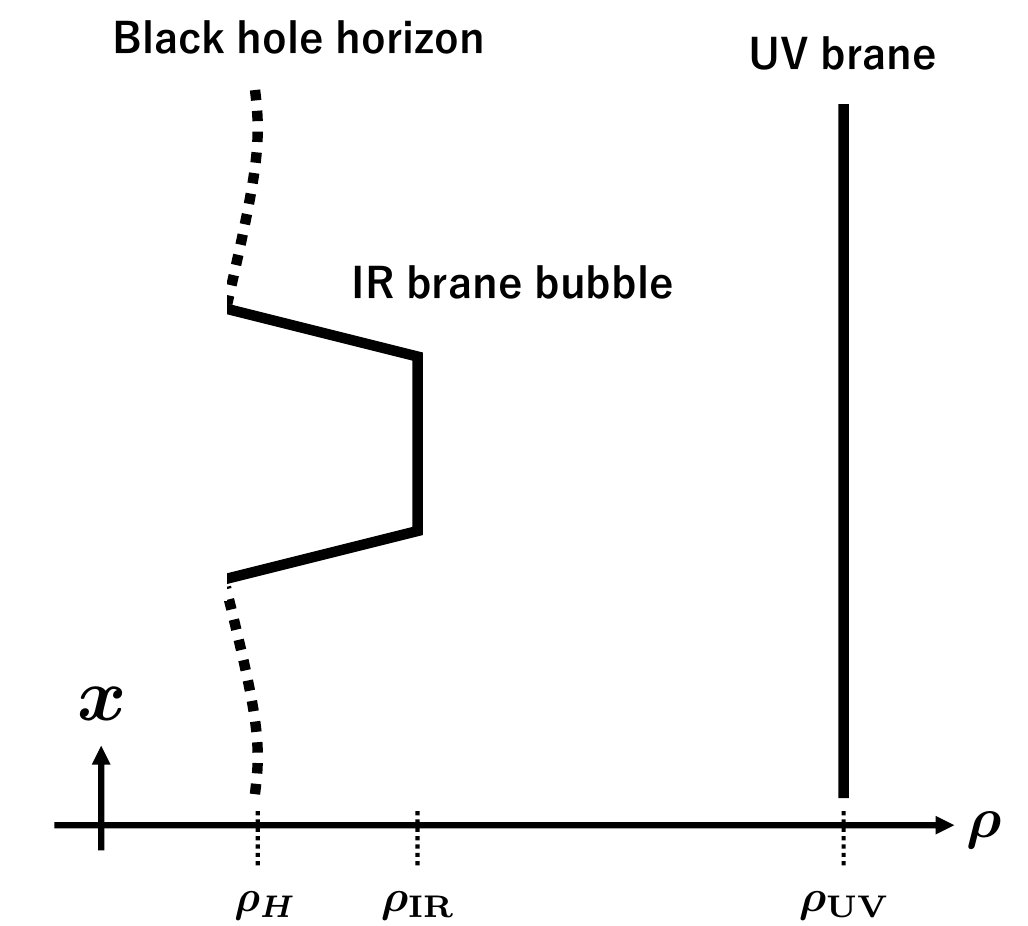}
\caption{
The schematic description of the IR brane bubble nucleation.
We here suppress two spatial dimensions and
$\rho_{\rm UV},~\rho_{\rm IR}$ and $T_H$ denote the positions of the UV and IR branes
and the event horizon of the AdS-S black hole.}
\label{fig:brane bubble}
\end{figure}
%%%%%%%%%%%%%%%%%%%%%%%%%%%%%%%%%%%%%%%%%%%%%%%%%%%%%%%%%%%%%

We next discuss the order of the phase transition and
how the phase transition between the RS spacetime and the AdS-S spacetime proceeds.
We have discussed the existence of two (local) minima of the free energies, at $T_H=T$ and $\mu=\mu_{\rm min}$, corresponding to the AdS-S spacetime and the RS spacetime.
Since the two minima are locally stable against fluctuations with respect to $T_H$ and $\mu$,
the phase transition occurs via the decay of the false vacuum.
Hence the order of the phase transition is expected to be of the first order.
The rate of the phase transition per unit volume per unit time is expressed as
\begin{align}
\Gamma =\mathcal{A}e^{-S} ,
\end{align}
where $\mathcal{A}$ is obtained by integrating out quantum (or thermal) fluctuations and
$S$ is estimated by solving the bounce equation in the semi-classical approximation.

In order to calculate $\Gamma$,
we should find a 5D gravitational instanton solution describing tunneling between the AdS-S spacetime and the RS spacetime.
The two solutions which we are interested in have different topologies: the AdS-S spacetime is simply connected while the RS spacetime is not.
As discussed in Ref.~\cite{Creminelli:2001th}, the AdS-S spacetime can be smoothly deformed into the RS spacetime by sending the event horizon to infinity ($T_H \to 0$) and back the IR brane from $\mu = 0$ through the AdS spacetime with the UV brane.
We assume that the 5D gravitational instanton solution is obtained by this deformation.
This is equivalent to the assumption that the relevant order parameter for the phase transition
in the RS spacetime 
is the radion field $\mu$ parametrizing the position of the IR brane 
while the one in the AdS-S spacetime is 
the Hawking temperature $T_H$ parametrizing the position of the black hole horizon. 
To maintain a valid effective field theory description of the RS spacetime,
the radion mass must be lighter than the mass of the first graviton KK-mode,
$m_{\rm radion} < m_{KK}\sim \pi \mu_{\rm min}$~\cite{Nardini:2007me}.
With this assumption, the phase transition proceeds via the ``IR brane bubble nucleation"
as schematically depicted in Fig.~\ref{fig:brane bubble}.
At high temperature, the system is in the AdS-S spacetime where the event horizon is placed at $T_H = T$.
As the temperature decreases, the event horizon moves toward $T_H = 0$.
Then, spherical brane patches on the horizon appear and they are eventually combined to form the IR brane.

\begin{figure}[t]
\centering\includegraphics[width=10cm]{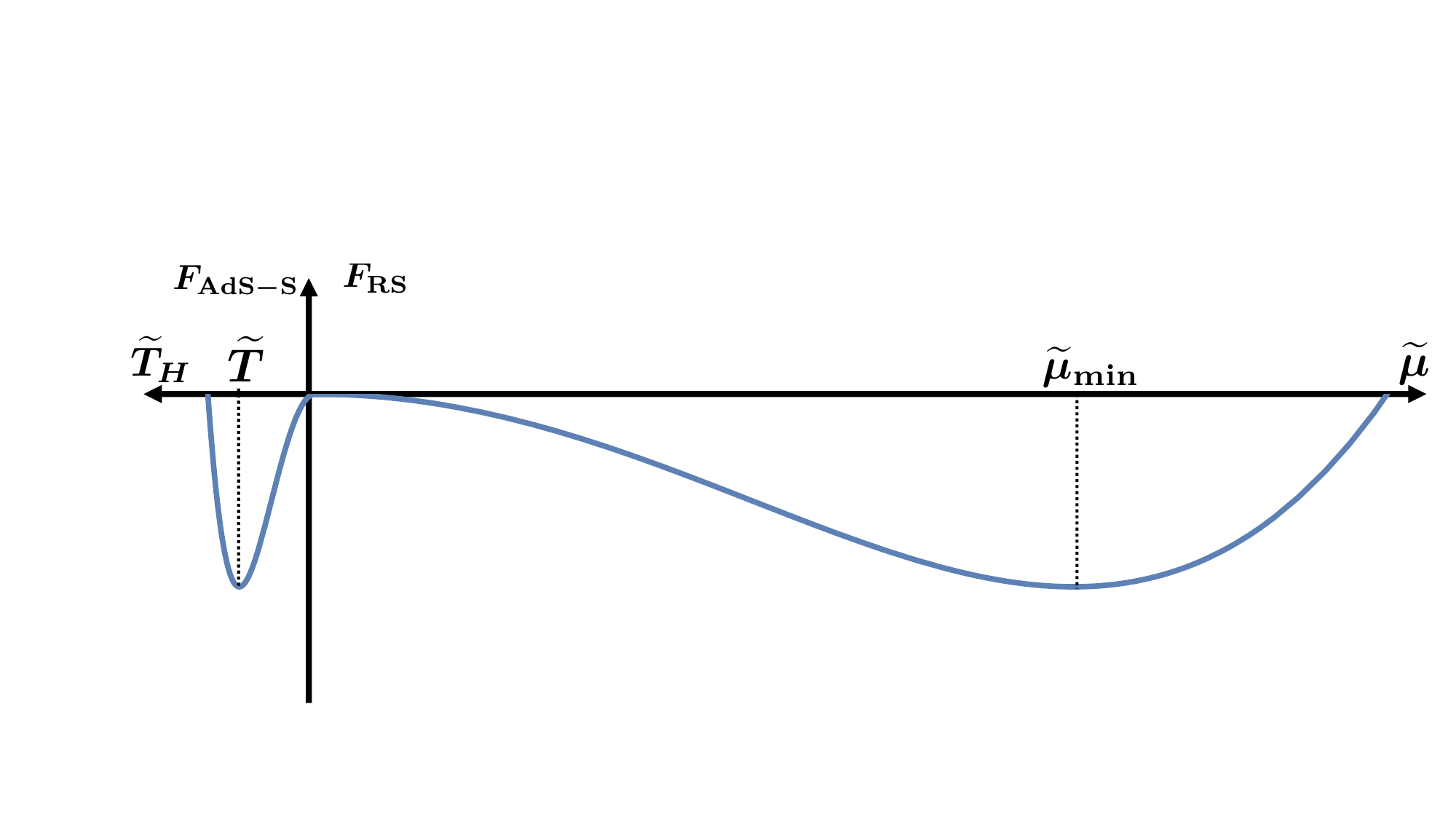}
\caption{The 4D potential of the AdS-S spacetime (left side of the axis) and the RS spacetime (right side of the axis) at the critical temperature $T_c$.
The canonically normalized Hawking temperature $\widetilde{T}_H(x)$ and the radion $\tilde{\mu}(x)$ are the order parameters in the AdS-S spacetime and the RS spacetime, respectively. 
The two origins $\tilde{\mu}=0$ and $\widetilde{T}_H=0$ must coincide with each other
because they both correspond to the bulk AdS spacetime with the UV brane.
}
\label{fig:free energies}
\end{figure}

When we consider the Hawking temperature as a spacetime dependent parameter, 
we can interpret $\Delta F_{\rm AdS-S}$ of Eq.~\eqref{eq:AdS-S free energy}
and $\Delta F_{\rm RS}$ of Eq.~\eqref{eq:RS free energy} as the 4D field theoretical potential for the $T_H(x)$ and $\mu(x)$ fields, respectively. 
However, we do not know the kinetic term for the Hawking temperature $T_H(x)$.
Since it is purely gravitational, we assume that the kinetic term is proportional to $N^2$
and take the form of $\frac{3N^2}{4\pi^2} c_1  \left( \partial T_H(x) \right)^2$ where $c_1$ is some $\mathcal{O}(1)$ coefficient
\cite{Creminelli:2001th,vonHarling:2017yew,Baratella:2018pxi}.
Fig.~\ref{fig:free energies} describes the potential of the Hawking temperature and radion fields after canonical normalization,
$\widetilde{T}_H(x)$ and $\tilde{\mu}(x)$.
Based on this potential, we can numerically calculate the tunneling rate 
from the AdS-S spacetime to the RS spacetime $\Gamma$ from the bounce action 
by using the standard under/over-shooting method.

%%%%%%%%%%%%%%%%%%%%%%%%%%%%%%%%%%%%%%%%%%%%%%%%%%%%%%%%%%%%%
%%%%%%%%%%%%%%%%%%%%%%%%%%%%%%%%%%%%%%%%%%%%%%%%%%%%%%%%%%%%%

Now we can briefly discuss that the potential in the regime of $\mu < T$ is not important 
to calculate the tunneling rate $\Gamma$~\cite{Creminelli:2001th}. 
We first note that $T_c \sim \mu_{\rm min} / \sqrt{N}$ from \eq{eq:critical temperature} with $V_{r, {\rm eff}} (\mu_{\rm min}) \sim \mu_{\rm min}^4$. This implies that $\mu_{\rm min} \gg T_c$ for a large $N$ 
and hence the potential (or free energy) around the minimum is justified. 
We also note that 
after the canonical normalization 
$\widetilde{\mu}$ is proportional to $N$ for a fixed $n$ 
while $\widetilde{T}_H$ is proportional to $N^{1/2}$ at the critical temperature.
Therefore, for a large $N$ the potential for $\widetilde{\mu}$ is very shallow while the potential for $\widetilde{T}_H$ is not that shallow (see Fig.~\ref{fig:free energies}).
Then the tunneling point is large, and hence, the gradient energy of the bubble is dominated
for a large $\tilde{\mu}$ where the 4D effective field approach is justified.
Throughout the analysis of the phase transition,
we consider this regime and calculate $\Gamma$ using the 4D effective theory.

\subsection{Transition rate}

We now calculate the bounce action to determine the parameter region where the phase transition
between the AdS-S spacetime and the RS spacetime is completed.
Generally, two types of bubble can be formed during the phase transition.
It was shown in Ref.~\cite{Linde:1981zj} that the bounce action at finite temperature is estimated as
\begin{align}
S= {\rm min} \left\{S_4 (T),~\frac{S_3 (T)}{T} \right\} ,
\end{align}
where $S_4 (T)$ and $S_3 (T)$ are the $O(4)$ and $O(3)$-symmetric bounce actions, respectively.
At low temperature, $T \ll T_c$, $S_4 (T)$ is less sensitive to $T$ and reaches a constant value due to the presence of the potential barrier at zero temperature, while $S_3 (T) / T$ has an explicit $T^{-1}$ enhancement.
We then find $S_4 (T) \ll S_3 (T) /T$ for $T \ll T_c$ and numerically confirmed this behavior.
We hence only consider the $O(4)$ symmetric bubble in the following discussion. 
It is calculated from 
\beq
 S_4 = \int dr 2 \pi^2 r^3 \lkk \frac{1}{2} \lmk \frac{d \phi}{d r} \rmk^2 + V(\phi) - V(\phi_f) \rkk, 
\eeq
where $\phi$ represents a canonically-normalized order parameter like $\widetilde{\mu}$ and $\widetilde{T}_H$, $V(\phi)$ is its potential (in the canonically-normalized basis), 
and $\phi_f$ is the order parameter at the false vacuum. 
The bounce solution $\phi(r)$ is determined by minimizing $S_4$ and should satisfy 
\beq
 \frac{d^2 \phi}{d r^2} + \frac{3}{r} \frac{d \phi}{d r} = V'(\phi). 
 \label{EoM}
\eeq
The boundary conditions are $d \phi (0)/ d r = 0$ and $\phi(r) \to 0$ for $r \to \infty$. This can be numerically computed by using the shooting method.

Although we numerically compute the bounce action, 
let us first estimate it semi-analytically by using a thick-wall approximation to understand its parameter dependence. 
For $T \ll T_c$, the bounce action is dominated by the gradient energy 
and thick-wall approximation gives a reliable estimate for $S_4$~\cite{Nardini:2007me}: 
\begin{align}
S_4 \simeq \frac{\pi^2}{2} \frac{\left.|\phi_t - \phi_f \right.|^4}{V(\phi_f) - V(\phi_t)} .\label{eq:thick-wall limit}
\end{align}
Here, $\phi_f$ and $\phi_t$ are field values at the false vacuum and tunneling point, respectively.
$\phi_t$ is determined by the requirement that the bounce action is minimized, i.e., by $\partial S_4/\partial \phi_t = 0$. 
In our situation, $\phi_f$ and $V(\phi_f)$ correspond to the Hawking temperature and  the free energy described by the AdS-S black hole \eqref{eq:AdS-S free energy}, respectively.
After canonically normalizing the kinetic terms of the Hawking temperature and radion fields,
we can write the bounce action \eqref{eq:thick-wall limit} as 
\begin{align}
S_4 \simeq \dfrac{9N^4}{8\pi^2} \frac{(\mu_t +\sqrt{c_1}T )^4}{V(\mu_{\rm min}) \left(\frac{T}{T_c}\right)^4 - V(\mu_t)}, \label{eq:bounce thick-wall}
\end{align}
where $\mu_t$ is the tunneling point.
This estimation shows that a shallower potential leads to a larger bounce action. 
The radion potential is very shallow in the basis where the radion kinetic term is canonically normalized, and thus, 
the tunneling rate is strongly suppressed for a large $N$. 
We note that 
$N$ must be larger than about $4.4$ from \eq{Nmin} 
so that the 5D Planck mass is larger than the AdS curvature in order to neglect the gravity loop corrections. 
This is one of the reasons that 
it is difficult to construct a radion stabilization mechanism in which 
the phase transition is completed fast enough.

Let us give a criteria for the transition rate, which must be fulfilled in order to avoid eternal inflation.
The phase transition can be completed only when the bubble nucleations are not diluted by the cosmic expansion.
This condition is given by $\Gamma > H^4$, where $H$ is the Hubble parameter.
At low-temperature, $T \ll T_c$, the energy density of the Universe is dominated by the vacuum energy of the radion potential.
Then the condition $\Gamma > H^4 $ can be written as follows,
\begin{align}
S_4 \leq 4 \log \left(\frac{m_{\rm radion} M_{\rm Pl}}{ \sqrt{F_{RS}}} \right) \simeq 140, \label{eq:criteria of the phase transition}
\end{align}
where we use $\mathcal{A} \sim m_{\rm radion}^4 = \mathcal{O}(1) \, {\rm TeV}^4$ and $H \sim \sqrt{\Delta F_{RS}}/M_{\rm Pl}$ during the vacuum domination.
We define the nucleation temperature $T_n$ as the temperature at which the phase transition is completed,
namely $S_4 (T_n) = 140$.

We here comment on 
the features of the phase transition
in the case of the Goldberger-Wise mechanism, where a radion potential comes from the energy of a bulk scalar field with mass $m_{\rm GW}$.
A parameter $\epsilon$ ($\equiv \sqrt{4+ m_{\rm GW}^2/k^2}-2$) 
is typically $\epsilon \sim 1/20$ to solve the hierarchy problem as noted in Ref.~\cite{Goldberger:1999uk}. 
In the limit of $\epsilon \to 0$, 
the radion is always stuck in the false vacuum, that is, $S_4\to \infty$, 
because the radion potential is scale invariant for $\epsilon = 0$ (or $m_{\rm GW} = 0$). 
In fact, as explicitly written in Ref.~\cite{Creminelli:2001th}, the $O(4)$ symmetric bounce action is proportional to $\epsilon^{-3/2}$. 
This factor leads to a strong suppression for the transition rate in the Goldberger-Wise mechanism.

One may note that the bounce action also depends on the vacuum expectation value of the Goldberger-Wise field on the IR brane~\cite{Creminelli:2001th}.
It is still possible to satisfy the condition \eqref{eq:criteria of the phase transition} with $N \gtrsim 4.4$ and $\epsilon\sim 1/20$ by making the vacuum expectation value large.
However, the large vacuum expectation value of the Goldberger-Wise field leads to a non-negligible back-reaction to the original RS spacetime, which is technically difficult to be taken into account.
As shown in Ref.~\cite{vonHarling:2017yew}, without taking into account the effects of the QCD confinement,
it is concluded that the parameter region which avoids eternal inflation has a non-negligible back-reaction to a pure gravity part and the analysis of the phase transition is unreliable in the Goldberger-Wise mechanism.

Now we shall turn to the analysis of the phase transition in our stabilization mechanism.
In our setup, we consider the phase transition between the AdS-S spacetime where the bulk $SU(N_H)$ gauge field is deconfined and the RS spacetime where the $SU(N_H)$ gauge field is confined.
One can evaluate the bounce action with the thick-wall approximation given in Eq.~\eqref{eq:bounce thick-wall}, where 
the tunneling point $\mu_t$ is determined by minimizing the bounce action 
in terms of $\mu_t$: 
\begin{align}
 \frac{\partial S_4}{\partial \mu_t} = 0.
\end{align}
In the limit of low $T$, 
the tunneling point $\mu_t$ is given by
\begin{align}
\mu_t = \mu_{\rm min} \left( \frac{1}{1-n}\right)^{\frac{1}{4n}}\left( \frac{2\lambda}{\gamma_c^4 n b_{\rm YM}}\right)^{\frac{1}{4(1-n)}}.
\end{align}
Substituting this into Eq.~\eqref{eq:bounce thick-wall}, 
we can estimate the bounce action. 

We also numerically solve \eq{EoM} by using the shooting method 
and compute the bounce action. 
The potential is given by the free energy; \eq{eq:AdS-S free energy} for $T_H$ and \eq{eq:RS free energy} for $\mu$. 
We note that the derivative of the potential for $\mu$ is not continuous at $\mu = \mu_c$ (see \eq{eq:radion potential}). 
To solve \eq{EoM} numerically, we continuously connect the gradient of the potential for $\mu > \mu_c$ and $\mu < \mu_c$ by using a hyperbolic tangent function, which 
looks similar to the Heaviside step function. 
From the numerical computation, 
we find that Eq.~\eqref{eq:bounce thick-wall} underestimates $S_4$ by a factor of $0.4$-$0.5$. 
We also find that the nucleation temperature $T_n$ is much higher than the QCD scale ($\sim 100 \MeV$), which fact is important for the gravitational wave production discussed in Sec.~\ref{sec:GW}.

\begin{figure}[t]
\includegraphics[width=7cm]{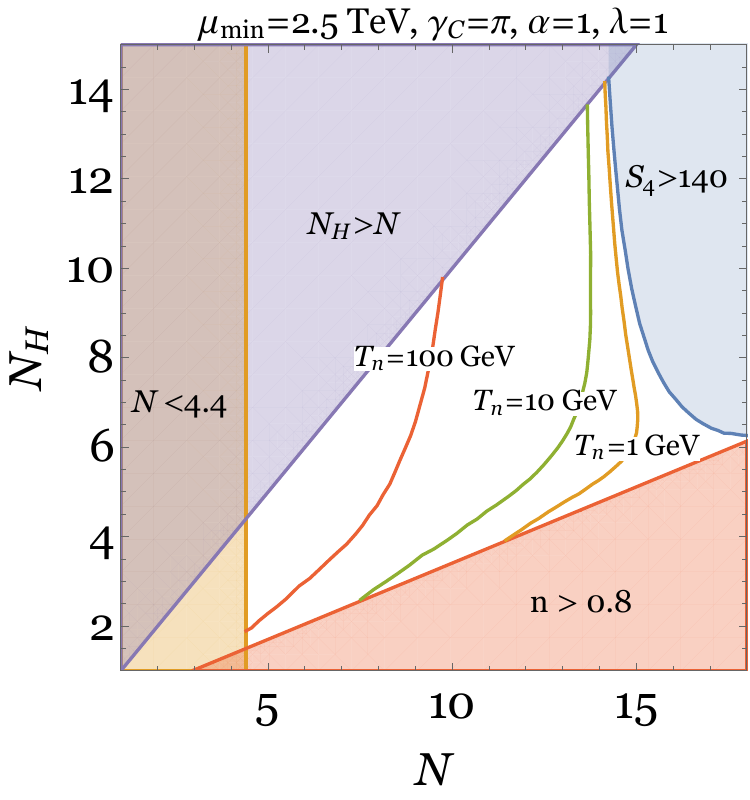}
\hspace{0.5cm}
\includegraphics[width=7cm]{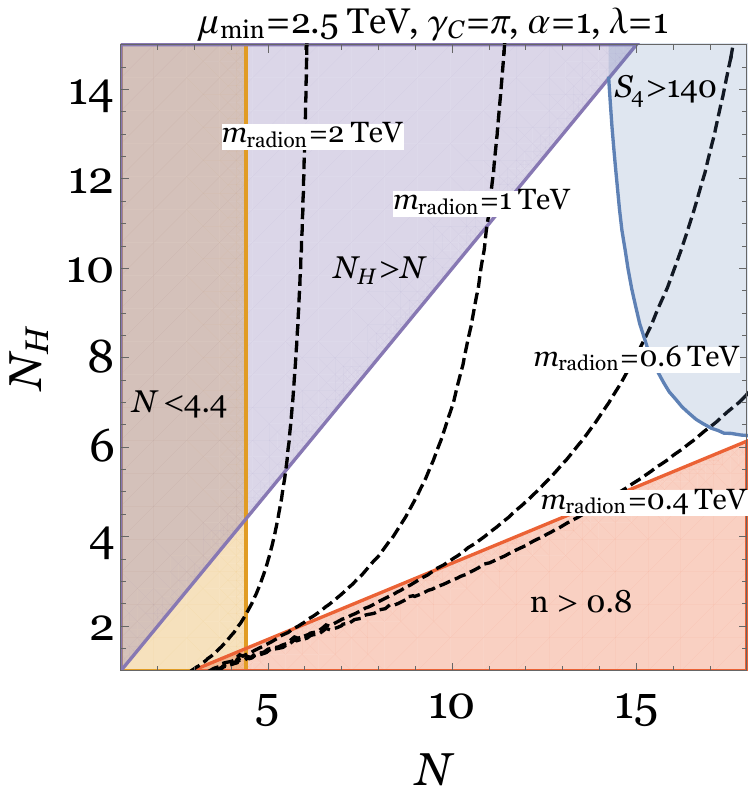}
\caption{
Exclusion plot in the $N_H$-$N$ plane. 
We take $\mu_{\rm min}=2.5 \, \rm TeV$,~$\gamma_c=\pi,~\alpha=1$, and $\lambda=1$. 
In the upper-right blue-shaded region, $S_4 > 140$ from our numerical calculations and the phase transition is not completed. 
In the bottom-right red-shaded region, $n > 0.8$ and $\mu_c$ is of order or smaller than the QCD scale, where the QCD effect has to be taken into account. 
In the upper-left blue-shaded region, $N_H > N$ and the finite temperature effect of $SU(N_H)$ has to be taken into account. 
In the left orange-shaded region, $N < 4.4$ and the quantum gravity effects have to be taken into account. 
The white region is allowed by those constraints. 
The contours in the allowed region in the left panel represent the nucleation temperature, 
while 
the dashed contours in the right panel represent the radion mass.
}
\label{fig:allowedregion}
\end{figure}

Figure~\ref{fig:allowedregion} shows the exclusion plot in the $N_H$-$N$ plane, 
where we take 
$\mu_{\rm min}=2.5 \,{\rm TeV},~\gamma_c=\pi,~\alpha=1$, and $\lambda=1$. 
The upper right corner is excluded by the criterion \eqref{eq:criteria of the phase transition} 
from our numerical calculation of the bounce action. 
We find that the phase transition can be completed for $N \lesssim 16$, which 
is large enough to be consistent with the condition \eq{Nmin}. 
We note that 
a finite-temperature effect of $SU(N_H)$, which is explained below \eq{eq:AdS-S free energy}, 
should be taken into account for $N_H > N$ (upper-left blue-shaded region). 
The nucleation temperature is presented by the contours in the allowed region in the left figure. 
One can see that it is typically of order $10$-$100$ GeV but can be as low as ${\cal O}(1) \GeV$. \footnote{
It is difficult to determine the precise nucleation temperature for the case of $\Lambda_H (\mu_c)<T_n$ 
because $SU(N_H)$ gauge theory is still in the deconfined phase for small $\mu$. 
We numerically confirm that $\Lambda_H (\mu_c) > T_n$ is satisfied in the whole allowed region in Fig.~\ref{fig:allowedregion} and hence our calculation is justified. 
}
The radion mass of Eq.~\eqref{eq:radion mass} is presented by the dashed contours in the right figure.

Here we comment on the bottom-right corner in Fig.~\ref{fig:allowedregion}, denoted as $n > 0.8$. 
It has been discussed in Ref.~\cite{vonHarling:2017yew} that the effect of the QCD modifies the radion potential for $\mu \sim \Lambda_{\rm QCD}$ ($\sim 100 \MeV$). 
This does not affect our calculation when $\mu_c \gtrsim \Lambda_{\rm QCD}$. 
However, $\mu_c$ may be as small as the QCD scale $\Lambda_{\rm QCD}$ 
for $0.8 \lesssim n < 1$ 
because the power of the parenthesis in \eq{muc} becomes very large for $n$ being close to unity. 
Thus we should take into account the effect of the QCD for $n \gtrsim 0.8$. 
For simplicity, we focus on $n \lesssim 0.8$ so that we can neglect its effect.

%%%%%%%%%%%%%%%%%%%%%%%%%%%%%%%%%%%%%%%%%%%%%%%%%%%%%%%%%%%%%
%%%%%%%%%%%%%%%%%%%%%%%%%%%%%%%%%%%%%%%%%%%%%%%%%%%%%%%%%%%%%

Fig.~\ref{fig:allowedregion} indicates that the phase transition is completed even for a relatively large $N$, 
where the gravity loop corrections are negligible. 
This is in contrast to the result in the Goldberger-Wise mechanism, where 
the phase transition is not completed for a large $N$. 
This fascinating result can be understood as follows. 
Our stabilization mechanism strongly breaks the scale invariance around $\mu = \mu_{\rm min}$ due to the confinement of the $SU(N_H)$ gauge interaction, 
while the Goldberger-Wise mechanism has a nearly scale invariant potential which is controlled by the small $\epsilon$ parameter.
As a result, the radion potential in our mechanism has a deeper minimum 
and the transition rate is larger than the one in the Goldberger-Wise mechanism with the same $N$. 
In addition, it should be noted that a back-reaction from the bulk hidden gauge field is trivially negligible
because the confinement scale is at the TeV scale which is very small compared to the 5D Planck mass. 
We also note that every dimensionless parameter in the model is of the order of unity and there is no fine tuning.

%%%%%%%%%%%%%%%%%%%%%%%%%%%%%%%%%%%%%%%%%%%%%%%%%%%
\section{Cosmological consequences} \label{sec:supercooled universe}
%%%%%%%%%%%%%%%%%%%%%%%%%%%%%%%%%%%%%%%%%%%%%%%%%%%

In this section, we discuss implications of the confinement-deconfinement phase transition on cosmology.
We estimate e-folding of inflation before the phase transition is completed and entropy injection that takes place after the transition.
In addition, we consider gravitational waves generated by the phase transition.
We also discuss production of $SU(N_H)$ glueballs.

\subsection{Entropy injection}\label{sec:entropy dilution}

As explicitly stated in Sec.~\ref{sec:phase transition}, our analysis of the phase transition by using the 4D effective theory description is only valid for $\mu_{\rm min}\gg T_c$.
Here we note that the energy density of the radiation and the vacuum energy of the radion at the critical temperature is roughly estimated as $\rho_{\rm rad}\sim T_c^4$ and $\rho_{\rm vac}\sim \Delta F_{\rm RS}\sim \mu_{\rm min}^4$, respectively. This implies that the vacuum energy dominates the energy density of the Universe and mini-inflation takes place before the phase transition is completed.

To be more precise, 
mini-inflation begins when the radiation energy becomes comparable to the vacuum energy: 
\beq
 \Delta F_{\rm AdS-S} - \Delta F_{\rm RS} = \frac{\pi^2}{90} g_*(T_{\rm inf}) T_{\rm inf}^4, 
\eeq
where $g_*(T)$ is the effective number of relativistic degrees of freedom 
and $T_{\rm inf}$ denotes the temperature at the beginning of mini-inflation. 
When $N$ is large, $\Delta F_{\rm AdS-S}$ is much larger than the radiation energy. 
We thus find that $T_{\rm inf} \simeq T_c$. 
The e-folding number of mini-inflation is then given by 
\begin{align}
 N_e \simeq \log \left(\frac{T_c}{T_n}\right).
\end{align}
From Fig.~\ref{fig:allowedregion}, 
we can see that the nucleation temperature $T_n$ is larger than of order $1 \GeV$ 
unless $N$ and $N_H$ are fine-tuned near the boundary of the blue-shaded region. 
Thus we find $N_e \lesssim  \log (1 \, {\rm TeV}/ 1 \, {\rm GeV}) \simeq 7$.

After the supercooled phase transition,
the free energy difference between the false vacuum and the true vacuum is injected into the RS phase.
We simply assume that the most of the free energy in the false vacuum $\Delta F_{\rm RS}$ is converted into the radiation in the true vacuum.
The reheating temperature, $T_{\rm RH}$, is then estimated as
\begin{align}
T_{\rm RH} \simeq \left(\frac{45}{4}\right)^{\frac{1}{4}} \frac{\sqrt{N}}{g_\ast^{\frac{1}{4}} (T_{\rm RH})}T_c ,
\end{align}
where we have used the definition of the critical temperature $T_c$ given by Eq.~\eqref{eq:critical temperature}.
Thus one can calculate the entropy injection after the strong first order phase transition from 
\begin{align}
\frac{s_n}{s_{\rm RH}} \simeq \frac{g_{*s} (T_n)}{g_{*s}(T_{\rm RH})} \left(\frac{T_n}{T_{\rm RH}}\right)^3, 
\end{align}
where $s_{\rm RH}$ and $s_n$ are entropy densities before and after the reheating 
and $g_{*s}$ is the effective number of relativistic degrees of freedom for entropy.

We briefly comment on cosmological consequences of the entropy injection. 
A late-time entropy production dilutes the relic abundance of the dark matter as well as the baryon asymmetry if they are produced before the phase transition (see, e.g., Refs.~\cite{Konstandin:2011dr,Servant:2014bla,Baratella:2018pxi}). 
In particular, 
the dilution factor is of order $10^{-9}$ for the case of $T_n / T_c = 10^{-3}$. 
One may therefore need a very large amount of dark matter and baryon asymmetry 
before the phase transition 
or 
need to produce them after the phase transition. 
We note that the latter possibility is not unlikely even if the nucleation temperature is as low as $1 \GeV$. 
For example, 
the QCD axion can be produced by the misalignment mechanism at the QCD phase transition, 
which takes place at $T \sim 0.1 \GeV$ and is not affected by the entropy dilution.\footnote{
This is the case when the Peccei-Quinn symmetry is spontaneously broken before 
the primordial inflation. 
For the case in which the Peccei-Quinn symmetry is spontaneously broken after the primordial inflation, 
see Ref.~\cite{Baratella:2018pxi}. 
}
Non-thermal production of weakly-interacting massive particles is also a viable scenario at a low-reheating temperature (see, e.g.,~\cite{Harigaya:2014waa,Harigaya:2019tzu}). 
The Affleck-Dine mechanism may be able to be realized to 
generate baryon asymmetry at a very low reheating temperature, 
by introducing a complex scalar field with a nonzero baryon charge~\cite{Affleck:1984fy, Dine:1995uk, Dine:1995kz}.
The cold electroweak baryogenesis scenario is also promising candidate for the generation of baryon asymmetry at a very low reheating temperature. (See e.g., Refs.~\cite{Krauss:1999ng,GarciaBellido:1999sv,Konstandin:2011ds})

\subsection{Generation of gravitational waves}
\label{sec:GW}

In this subsection, we consider GW signals produced during or after the phase transition.
The amplitude and frequency of GW signals generated by a first-order phase transition mainly depend on two parameters called duration and latent heat density.
The duration of the phase transition denoted by $\beta$ is defined as the time variation of the nucleation rate of bubbles~\cite{Caprini:2015zlo}:
\beq
\frac{\beta}{H(T_{\rm RH})} 
&\equiv&\frac{1}{H(T_{\rm RH}) \Gamma} \frac{d \Gamma}{d t}
\\
&\simeq& 
\frac{H (T_n)}{H(T_{\rm RH})}  T_n \left. \frac{d S_4}{dT} \right|_{T=T_n}.
\eeq
This is typically ${\cal O}(1$-$100)$ for the case we are interested in. 
The latent heat density normalized by the energy density of the radiation $\rho_{\rm rad}$ is approximately given by~\cite{Caprini:2015zlo}
\begin{align}
\alpha \simeq \frac{\abs{\Delta F_{\rm RS}}}{\rho_{\rm rad}(T_n)},
\end{align}
where we used $\abs{\Delta F_{\rm AdS-S}} \ll \abs{\Delta F_{\rm RS}}$ at the nucleation temperature. 
Since the vacuum energy usually dominates before the phase transition is completed, 
we find $\alpha \gg 1$ in our model.

The GW amplitude $\Omega_{\rm GW}h^2$ is decomposed into three different contributions,
\begin{align}
\Omega_{\rm GW} h^2 = \Omega_{\rm col}h^2 + \Omega_{\rm sw}h^2+ \Omega_{\rm turb}h^2,
\end{align}
where $\Omega_{\rm col}h^2,~\Omega_{\rm sw}h^2$ and $\Omega_{\rm turb}h^2$ denote the contributions from bubble collisions~\cite{Turner:1990rc,Kosowsky:1991ua,Kosowsky:1992vn,Turner:1992tz,Jinno:2016vai,Jinno:2017fby}, sound wave~\cite{Hindmarsh:2013xza,Giblin:2014qia,Hindmarsh:2015qta,Hindmarsh:2017gnf}, and turbulence~\cite{Kamionkowski:1993fg,Caprini:2006jb,Caprini:2009yp,Kosowsky:2001xp,Gogoberidze:2007an,Niksa:2018ofa} of the thermal plasma, respectively.
In the absence of the thermal plasma, most of the released energy is converted into the kinetic energy of the accelerating bubble wall, and hence, the bubble wall velocity before collisions is very close to the speed of light.
This bubble is called {\it runaway} bubble~\cite{Caprini:2015zlo}.
When the runaway bubble is realized, bubble collisions give a dominant contribution to the total GW signals.
On the other hand, when the thermal plasma is presented, the accelerating bubble wall receives a friction from the thermal bath~\cite{Bodeker:2009qy}.
If the bubble wall velocity $v_w$ reaches a terminal velocity 
due to the friction, most of the kinetic energy of the accelerating bubble wall is injected into the thermal bath.
In this case, sound wave and turbulence of the plasma become the main source of GW signals.

In the case of the electroweak phase transition, a friction emitting the electroweak gauge bosons called {\it transition radiation} gives
a significant contribution to the force acting on the bubble wall~\cite{Bodeker:2017cim}. 
In our case, there are $SU(N_H)$ gauge fields that 
are in the thermal plasma outside the bubble 
and are strongly interacting (and are confined) inside the bubble. 
Although the $SU(N_H)$ gauge fields are strongly interacting inside the bubble and
the bubble dynamics in the present setup differs from the case of the electroweak phase transition, 
we still expect that the transition radiation occurs and gives an important contribution to the friction force. 

Suppose first that we can neglect the friction effect. 
Then walls accelerate due to the pressure of the vacuum energy 
until they collide. 
The Lorentz gamma factor at the time of collision, $\gamma_{*}$, 
is roughly given by~\cite{Ellis:2019oqb}
\begin{align}
\gamma_* \sim \frac{R_*}{R_0} ,
\end{align}
where $R_*$ and $R_0$ are the bubble radius at the time of collision and the formation of bubble, respectively. 
Since the typical distance among bubbles is of order $\beta^{-1}$, we can estimate $R_* \sim \beta^{-1}$. 
The initial bubble radius is determined by the instanton solution 
and is determined by the curvature of the radion potential, namely $m_{\rm radion}^{-1}$. 
Here we note that the transition takes place via the O(4) bounce action 
rather than the O(3) bounce action in our model.

However, the wall velocity cannot be arbitrary large because of the friction effect. 
The pressure acting on the bubble wall is eventually balanced between the vacuum energy and the friction 
due to the transition radiation. 
The Lorentz gamma factor of the bubble wall at the terminal velocity, $\gamma_{\rm eq}$, 
is roughly estimated as~\cite{Baratella:2018pxi}
\begin{align}
\gamma_{\rm eq} \sim \frac{\abs{\Delta F_{\rm RS}}}{ g^2 \Delta m_V T_n^3} 
\sim 
\left(\frac{T_c}{T_n}\right)^3,
\end{align}
where $g$ is the gauge coupling constant of $SU(N_H)$ and $\Delta m_V$ is the mass difference of the gauge boson inside and outside the bubble. We assume $g \sim 1$ and $\Delta m_V \sim \Lambda_{H,0}$ ($\sim T_c$) though the precise values are not relevant for the resulting gravitational waves. 
When $\gamma_*$ exceeds $\gamma_{\rm eq}$, bubble walls reach the terminal velocity before they collide. 
In this case, sound waves and turbulence of the plasma are the main source of GWs. 
This condition turns out to be
\begin{align}
T_n \gtrsim 10^{-2} \, {\rm GeV}  \times \left( \frac{\beta}{H(T_{\rm RH})} \right)^{1/3} 
\lmk \frac{T_c}{1 \TeV} \rmk^{4/3}.
\label{eq:nucleation criteria}
\end{align}
Noting that $\beta / H(T_{\rm RH}) = {\cal O}(1$-$100)$, 
we find that this is usually satisfied in our case (see Fig.~\ref{fig:allowedregion}). 
Thus we calculate GW signals sourced by sound waves and turbulence of the plasma below.

The contribution to the GW amplitude from sound waves, $\Omega_{\rm sw} h^2$, is given by~\cite{Hindmarsh:2015qta}.\footnote{It was pointed out in Ref.~\cite{Ellis:2018mja} that this formula overestimates GW signals. 
One may regard \eq{eq:sound wave} as an upper bound for GW signals. 
%Here we give an upper bound for GW signals.
}
\begin{align}
\Omega_{\rm sw}(f) h^2 \simeq 2.65\times10^{-6} \times \left(\frac{H(T_{\rm RH})}{\beta} \right)\left(\frac{\kappa_{\rm sw}\alpha}{1+\alpha} \right)^2 \left(\frac{100}{g_*}\right)^{\frac{1}{3}} v_w \left(\frac{f}{f_{\rm sw}}\right)^3 \left( \dfrac{7}{4+3\left(\frac{f}{f_{\rm sw}}\right)^2}\right)^{\frac{7}{2}}, \label{eq:sound wave}
\end{align}
where $\kappa_{\rm sw}$ and $v_w$ are the efficiency factor and the bubble wall velocity, respectively.
In our case, we simply set these values as $\kappa_{\rm sw}\simeq 1$ and $v_w \simeq 1$ because of the strong supercooled phase transition, $\alpha \gg1 $.
The peak frequency $f_{\rm sw}$ is roughly given by $2 \beta / \sqrt{3} v_w$ 
with a redshift factor: 
\begin{align}
f_{\rm sw} \simeq  1.9\times 10^{-4} \, {\rm Hz} \times \frac{1}{v_w} \left(\frac{\beta}{H(T_{\rm RH})}\right) \left(\frac{T_{\rm RH}}{1 \, {\rm TeV}}\right) \left( \frac{g_*}{100}\right)^{\frac{1}{6}} .
\end{align}
On the other hand, the contribution to the GW amplitude from turbulence, $\Omega_{\rm turb} h^2$, is given by~\cite{Caprini:2009yp,Binetruy:2012ze}
\begin{align}
\Omega_{\rm turb}(f) h^2 \simeq  3.35 \times 10^{-4} \times \left(\frac{H(T_{\rm RH})}{\beta}\right) \left(\frac{\kappa_{\rm turb}\alpha}{1+\alpha}\right)^{\frac{3}{2}} \left(\frac{100}{g_*}\right)^{\frac{1}{3}} v_w \dfrac{\left(\frac{f}{f_{\rm turb}}\right)^3}{\left(1+\frac{f}{f_{\rm turb}}\right)^{\frac{11}{3}}\left(1+\frac{8\pi f}{h_*}\right)} ,
\end{align}
where $h_*$ is given by 
\begin{align}
h_* \simeq 1.65 \times 10^{-4} \, {\rm Hz} \times \left(\frac{T_{\rm RH}}{1 \, {\rm TeV}}\right)\left(\frac{g_*}{100}\right)^{\frac{1}{6}}.
\end{align}
The peak frequency $f_{\rm turb}$ is roughly given by $3.5 \beta / 2 v_w$ 
with a redshift factor: 
\begin{align}
f_{\rm turb} \simeq 2.7 \times 10^{-4} \, {\rm Hz} \times \frac{1}{v_w}\left(\frac{\beta}{H\left(T_{\rm RH} \right)}\right) \left(\frac{T_{\rm RH}}{1 {\rm TeV}}\right) \left( \frac{g_*}{100}\right)^{\frac{1}{6}}.
\end{align}
The fraction of latent heat that is transformed into turbulence, $\kappa_{\rm turb}$, 
is assumed to be $\kappa_{\rm turb}=0.05 \kappa_{\rm sw}$ for a conservative estimation~\cite{Caprini:2015zlo}. 

%%%%%%%%%%%%%%%%%%%%%%%%%%%%%%%%%%%%%%%%%%%%%%%%%%%%%%%%%%%%%
\begin{figure}[t]
\centering\includegraphics[width=10cm]{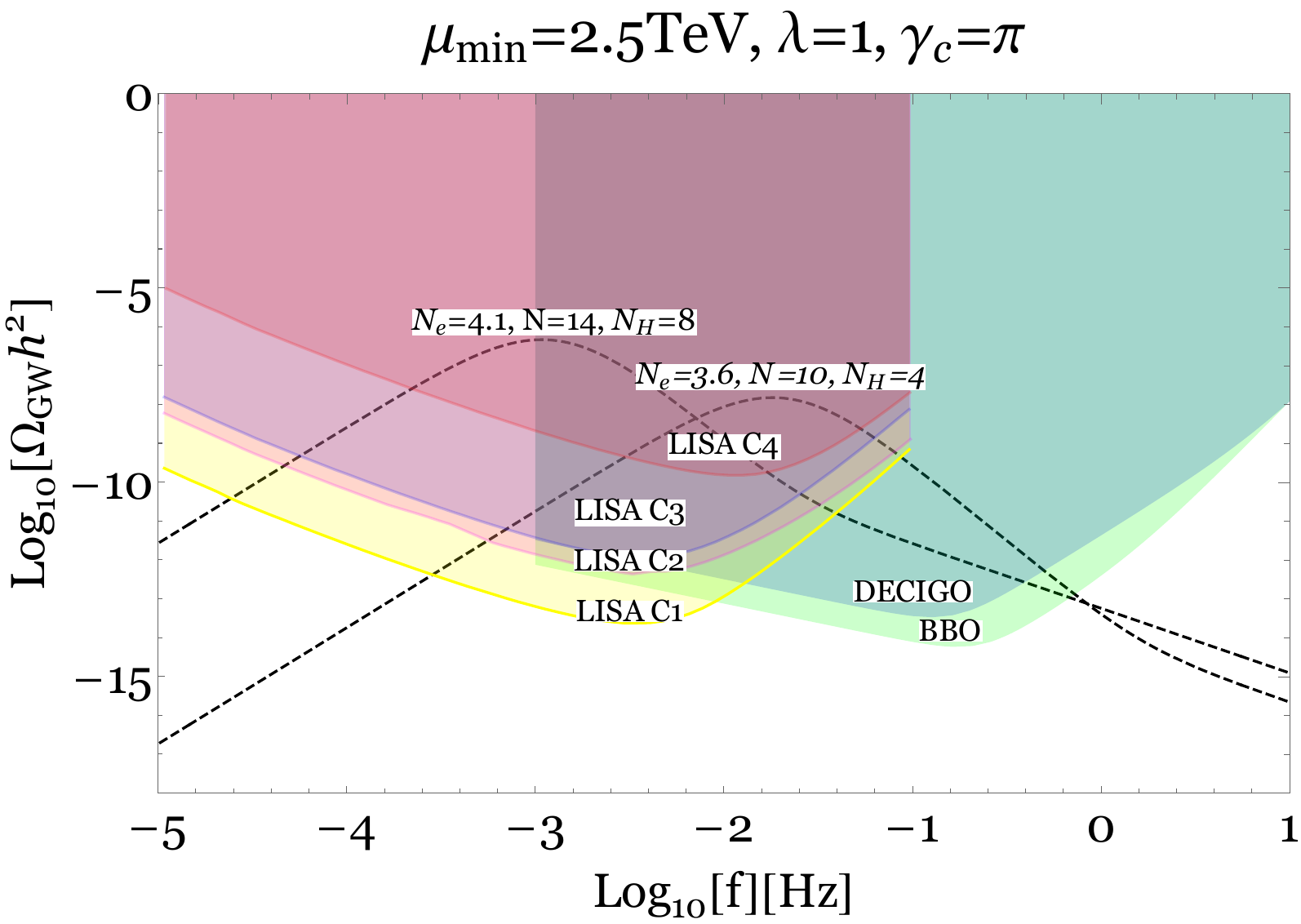}
\caption{
The GW amplitude generated from the phase transition. Our benchmark points are $\mu_{\rm min}=2.5 \, \rm TeV$, $\lambda = 1, \gamma_c=\pi$, $N=14,~N_H=8$ ($N_e= 4.1$) and $N=8,~N_H=3$ ($N_e =3.8$), respectively.
Detectable regions by eLISA~\cite{Klein:2015hvg,Caprini:2015zlo},~DECIGO and BBO~\cite{Yagi:2011wg} are also shown.
}
\label{fig:GW}
\end{figure}
%%%%%%%%%%%%%%%%%%%%%%%%%%%%%%%%%%%%%%%%%%%%%%%%%%%%%%%%%%%%%

We plot the GW signals in Fig.~\ref{fig:GW}. Our benchmark points are $\mu_{\rm min}=2.5 \, \rm TeV$, $\lambda = 1, \gamma_c=\pi$, $N=14,~N_H=8$ ($T_n\simeq 7.8 \, {\rm GeV},~\beta / H(T_{\rm RH})\simeq 5.7,~N_e\simeq 4.1$) and $N=8,~N_H=3$ ($T_n\simeq 10 \, {\rm GeV},~\beta / H(T_{\rm RH})\simeq 124,~N_e\simeq 3.8$), respectively.
We can see from the figure that the GW signals reach a detectable region by LISA, DECIGO and BBO.
We therefore find that our model can be probed by the detection of GW signals. 
However, in order to give a reliable estimate for the GW signals, we need to clarify the bubble dynamics, including a thermal friction coming from the $SU(N_H)$ gauge interaction. 
It should be also noted that the validity of the formula \eqref{eq:sound wave} is unclear
for the strong supercooled phase transition, $\alpha \gg1$
(see Ref.~\cite{Jinno:2019jhi} for a recent study).
However, these issues are beyond the scope of the present paper and should be discussed elsewhere.

\subsection{Glueball production}

In the RS spacetime, the $SU(N_H)$ gauge field is confined and then
the corresponding glueballs may be formed.
The lightest state of the glueballs is a CP-even scalar state $0^{++}$
and its mass is estimated by the lattice calculation as $m_{0^{++}} \approx 7\Lambda_{H,0}$ (see e.g. \cite{Nakai:2015swg}
for the summary of the spectrum).
This glueball is produced after the phase transition only when $T_{\rm RH} \gtrsim m_{0^{++}}$ is satisfied.
We have found that this condition is never satisfied for $N\gtrsim 4.4$.
Thus, the glueball $0^{++}$ does not lead to any cosmological concern.
Note that the radion mass is lighter than the glueball mass for the interesting parameter space,
and hence, the radion does not decay into the glueballs.
The second, third, and fourth lightest states are $2^{++}$ with mass $m_{2^{++}} \approx 10\Lambda_{H,0}$,
$0^{-+}$ with mass $m_{0^{-+}} \approx 11\Lambda_{H,0}$, 
and $1^{+-}$ with mass $m_{1^{+-}} \approx 12\Lambda_{H,0}$, respectively. 
Among them, the $0^{-+}$ state is stable if CP is not broken in this sector.
We can let this state decay by introducing a nonzero theta term in the $SU(N_H)$ gauge theory, 
which is rather natural unless the exact CP invariance is assumed. 
The same is true for the other CP-odd states and then they do not cause a cosmological problem even if they are produced
by some mechanism.

\section{Conclusion} \label{sec:conclusion}

In this paper, we have proposed a new radion stabilization mechanism in the RS model and
investigated dynamics of the phase transition between the AdS-S spacetime and the RS spacetime.
We introduced a bulk $SU(N_H)$ gauge field which confines at a TeV scale.
This condensation generates a Casimir energy which contributes to the radion potential negatively.
We assume that the IR brane tension is deviated from the value used in the original RS spacetime. 
Then, the radion potential can be stabilized by the balance between the Casimir energy and the brane tension.
It turns out that the radion potential has a local minimum at the origin and the global minimum at a TeV scale, similar to the radion potential generated by the Goldberger-Wise mechanism. 
The TeV scale arises due to the strong dynamical effect of the $SU(N_H)$ gauge theory, 
so that it is natural due to the dimensional transmutation.

When the radion stabilization mechanism is presented, the RS spacetime is energetically favored below the critical temperature
which is typically at a TeV scale.
We saw that the phase transition from the AdS-S spacetime
is the first order phase transition and proceeds via the IR-brane bubble nucleation.
By the detailed analysis, it was found that the phase transition takes place via a supercooling phase and can be completed even for $N\gtrsim 4.4$, in which gravity loop corrections are suppressed.
Since the confinement scale is at a TeV scale, a back-reaction due to the introduction of the hidden $SU(N_H)$ gauge field to the original RS spacetime is trivially negligible.
We compared our result to that obtained by the Goldberger-Wise mechanism and showed that
in our model the phase transition is completed faster than the case of the Goldberger-Wise without any problems. 

We determined the nucleation temperature, which is typically of order $10$-$100 \GeV$. 
If it is low enough, a mini-inflation occurs before the phase transition is completed. 
Since the entropy is generated from the vacuum energy, 
dark matter abundance and baryon asymmetry are diluted after the phase transition. 
We also estimated the GW spectrum generated by the phase transition 
and found that it can be detected by future experiments such as eLISA, DECIGO and BBO.
The detection of such GW signals will be one of the important probes of our model.

\section*{Acknowledgments}

KF was supported by JSPS and NRF under the Japan-Korea Basic Scientific Cooperation Program and would like to thank participants attending the JSPS and NRF conference for useful comments.
YN would like to thank Rutgers University where he belonged to when the present work was initiated.
MY was supported by JSPS Overseas Research Fellowships and the Department of Physics at MIT. 
MY was also supported by the U.S. Department of Energy, Office of Science, Office of High Energy Physics of U.S. Department of Energy under grant Contract Number DE-SC0012567.

%\appendix

\bibliography{ref}

\end{document}